\documentclass{icrc2009}

\usepackage{graphicx}   % for including figures
\usepackage{caption}   % for captions
\usepackage{fixltx2e}
\usepackage{url}

\newcommand{\shorttitle}[1]%
{\markboth{Proceedings of the 31\MakeLowercase{$^{st}$} ICRC, {\L}\'{o}d\'{z} 2009}{#1} }
\newcommand{\etal}{\MakeLowercase{\textit{et al. }}} % "et al."

\begin{document}
\title{Contribution of Stellar Flares to the Diffuse Component of Galactic Gamma-Rays
}

\author{\IEEEauthorblockN{{Y. Muraki }\IEEEauthorrefmark{1}}
        
\IEEEauthorblockA{\IEEEauthorrefmark{1}Department of Physics, Faculty of Science, Konan University, Kobe 658-8501, Japan}}

% please write the preseter's name and short title (3-4 words maximum)
%    which will appear at the header of the even pages.
\shorttitle{Muraki \etal Origin of CR and diffuse gamma-rays}
\maketitle

\begin{abstract}It is well known that diffuse gamma-rays are produced by collisions between Galactic cosmic rays and interstellar matter (Hayakawa$-$Hutchinson$-$Morrison hypothesis). We estimate the contribution of high energy protons produced by stellar flares to the diffuse component of Galactic gamma-rays. Each star in the Galaxy regularly produces flares that generate high energy protons in the energy range from 100 MeV to 1 TeV. These protons are a likely source of cosmic rays. We estimate that the Galactic bulge can constantly generate $2.7\times 10^{14}$ A of high energy proton current.

\end{abstract}

 
\section{Introduction}
 According to the Hayakawa$-$Hutchinson$-$Morrison hypothesis, the diffuse component of Galactic gamma-rays is produced by the collision of high energy cosmic-rays with interstellar matter [1-3], while according to long-held theories, super novae [4] or the interstellar medium [5] are possible sources of cosmic rays [6] (See Figre 1).   In this paper, we show the other possible source of cosmic rays; the diffuse component of Galactic gamma-rays originate from the flare particles of stars throughout the Galaxy.
 
As an analogy for gamma-rays, we consider the model of solar wind. According to current theory, solar wind is formed by the cumulative effect of micro and nano flares on the solar surface [7-8]. The duration of these micro and nano flares is short and their scale is small. However, these small solar flares occur very frequently and hence a continuous high speed solar wind is maintained with a constant intensity [9-10].

With reference to this model, we propose that the origin of the diffuse component of Galactic gamma-rays may be attributed to continuously produced high energy protons originating from stellar flares. In the Galactic bulge, there are about $1.6\times 10^{10}$ stars. (We call this number {\bf  $\sharp$A} ).   
Therefore, although stellar flares are small, they occur frequently in each star, and hence their emissions create a steady high energy proton current with constant intensity. Thus, the bulge of our Galaxy can be likened to a powerful cosmogenic DC current generator.

Occasionally a star in the bulge releases a large flare, emitting high energy particles beyond {\it E}p $\geq$100 GeV.  These high energy particles collide with the stellar atmosphere or interstellar matter and produce gamma-rays with energy $E_{\gamma}{\geq}$10 GeV.   In the next section, we show numerically that the present model can explain the intensity of the diffuse component of Galactic gamma-rays.

 \section{Fundamental Data used for Calculations}

Before describing our model numerically, we show here several numerical values that are necessary for defining the model.

(1) It is well known that the Sun is an exceedingly typical star in our Galaxy. Therefore, we assume that the frequency of stellar flares throughout the Galaxy can be represented by that of solar flares, i.e. the frequency of flares on the Sun.

Solar flares are classified into A, B, C, M and X classes, by size, as identified by X-ray observations [11]. They are represented by a logarithmic scale such that the difference in intensity between classes A1 and X10 is a factor of 100,000.   The acceleration of particles over 100 GeV is usually observed for the largest classes of solar flare, such as class X10 [12]. It has been reported that the frequency of solar flares is inversely proportional to the size of the flare, and the product of the frequency and size is constant [13]. Based on this estimation, a flare of size X10 occurs every 3$\times10^{7 }$ seconds (once a year), while flares of size M1 occur every 3 days on average.

(2) The energy spectrum of the accelerated protons for each star can be represented by an example observed solar flare of April 15, 2001 (the Easter event). In this flare the energy spectrum of the accelerated protons can be expressed by a power law with gamma index $\Gamma$ = -3.75 (differential index) in the energy range between {\it E}p = 200 MeV and 100 GeV [12].  The proton fluxes at the Earth for {\it E}p $>$100 MeV, $>$1 GeV and $>$10 GeV are 100, 1 and $10^{-3}$ protons/(${\rm cm}^{2}$ sec str), respectively. (We call these numbers 
 {\bf  $\sharp$B}).

(3)In this paper the energy spectrum of photons induced by stellar flares is represented by a similar power spectrum for the photons in the energy range $E_{\gamma}{\approx}17$\,MeV to 8\,GeV with spectral index  ${\Gamma} = -3.75$.   Here we have assumed the inelasticity of gamma-rays to be $K_{\gamma}{\approx}1/12$.
   
A detailed Monte Carlo calculation is necessary on the secondary production process; however, the gross features of the interactions are not changed dramatically from this estimation. Therefore, the intensity of gamma-rays at $E_{\gamma}> 75$ MeV may be described by the proton intensity at {\it E}p $\geq$900 MeV $\sim$ 1 GeV.
It must be noted here that pions will be not produced by protons with energy less than {\it E}p$\leq$ 1 GeV.
(See note in [18].)

(4) In the largest solar flares, the peak intensity of accelerated protons occurs during a period of 1200 seconds on average. Therefore if 2.5$\times 10^{4 }$ stars out of a total 1.6$\times 10^{10}$ stars in the Galactic bulge produce one X10 class stellar flare once every year, high energy protons are constantly produced in the bulge on average (1200 sec$\times $2.5$\times $$10^{4 }$= 3$\times$$10^{7 }$sec = 1 year).    Hence, the bulge is likely to be the main source for high energy protons of {\it E}p $\geq$1 GeV. 

  Dividing the total number of stars in the bulge by the number of flare stars with intensity  X10 per year, we obtain 1.6$\times 10^{10 }$stars/2.5$\times 10^{4 }$stars = $ 6.4\times10^{5 }$ (we call this number  {\bf  $\sharp$C}), which represents the number of stars in the Galactic bulge that may generate a constant high energy proton current with energy {\it E}p $\geq$ 1 GeV.
 
(5) To calculate the total high energy proton flux induced by solar flares we use the total number of protons accelerated from the flare of April 15, 2001 as a standard value. The observed flux of accelerated protons for the Easter flare was observed to be ${\sim}1 {\rm proton}/({\rm cm}^{2} {\rm sec str})$ at Earth for {\it E}p $\geq $1 GeV (from  {\bf  $\sharp$B}). The total flux at the Sun of the Easter flare is derived by multiplying the distance correction factor between the Sun and the Earth, i.e., $4{\pi}r^{2} = 2.0{\times}10^{23}{\rm m}^{2}= 2.7{\times}10^{27}{\rm cm}^{2}{\rm str}$, where r corresponds to 1 AU = $1.5\times 10^{11}$ m.   In other words, if protons were emitted in an isotropic direction from the Sun, the high energy proton flux with Ep $\geq $1 GeV (for which  {\bf  $\sharp$B}$\sim$ 1) at the Sun is 2.7$\times 10^{27}$ protons/sec. This value corresponds to a proton current of 4.3$\times 10^{8}$A with Ep $\geq $1 GeV in the flare. (We call this number {\bf  $\sharp$D})

\section{Calculations and Results}

 In this section, we obtain the expected flux of protons accelerated by stellar flares in the bulge and compare them with observed data.

(1) The total high energy proton intensity produced by stellar flares in the bulge is determined by estimating the total proton current induced at the bulge using  {\bf  $\sharp$C}. We assume that each star in the bulge releases a large stellar flare once every year, creating a strong intensity that continues for about 1200 seconds. We further assume that the flux of high energy protons generated by the flare can be approximated by the Easter flare. Then, the total flux induced by the stellar flares in the bulge can be represented by  {\bf  $\sharp$C}$\times${\bf  $\sharp$D} [(number of effective stars) 
$\times$(total number of protons produced by each star)] or (6.4$\times10^{5}$) $\times$(2.7$\times10^{27}$) = 1.7$\times10^{33}$ protons/sec for {\it E}p$\geq$1 GeV. This corresponds to {\bf 2.7$\times10^{14 }$A} and indicates that the Galactic bulge constantly generates a very strong DC current.

(2) When calculating the total number of high energy protons in the Galaxy it is important to consider the Galactic magnetic field. If there was no magnetic field in the Galaxy, the generated protons, generated at a rate of 1.7$\times10^{33}$ protons/sec, would be leaving the galaxy, dissipating into intergalactic space. However, the galactic magnetic field works to retain the generated protons in the Galaxy for a long time. From the measurement of $Be^{10}$ we estimate this time to be 20 million years [14]. Therefore, the total number of high energy protons in the Galaxy is 1.7$\times10^{33}$$\times$ 2.0$\times10^{7}$ = 3.4$\times10^{40}$ protons. (We call this number  {\bf  $\sharp$E}.)

(3) As the produced protons travel through space, they undergo nuclear interactions with interstellar gas, as pointed out by Hayakawa, Hutchinson and Morrison. The probability of a nuclear interaction occurring can be calculated by assuming a total cross-section of 40 mb for protons and a gas density of one hydrogen atom/${\rm cm}^{3}$.  For a path of 1 kpc (= 3${\times}10^{21}$cm), the collision rate is 40$\times10^{-27}$ [${\rm cm}^{2}$] $\times$3$\times10^{21}$ [cm] $\times$ 1 [/${\rm cm}^{3}$] = 1.2$\times10^{-4}$.   Over 20 million years, protons with {\it E}p$\geq$1 GeV traveling at about light speed travel a distance of about 6.1${\times}10^{3}$kpc, giving a product of (6.1${\times}10^{3}$) $\times$ (1.2${\times}10^{-4}$) = 0.73. Hence, about 73$\%$of produced protons lose their energy by nuclear interactions with intermediate gas. 
(We call this number {\bf  $\sharp$F}).

(4) The production rate of gamma-rays $E_{\gamma}{\geq}$70 MeV in the Galaxy is estimated by 
 {\bf  $\sharp$E}$\times${\bf  $\sharp$F}= 3.4${\times}10^{40 }{\rm protons} {\times}$0.73 = 
2.5$\times10^{40}$ photons/sec. (We call this number  {\bf  $\sharp$G}.)

(5) For the observation of the Galactic bulge by gamma-ray telescopes, we estimate the geometrical acceptance of the bulge by detectors on the Earth. The Galactic center is located about 8 kpc from the Earth. The intensity of the gamma-rays emitted at the center of the Bulge is reduced by the geometrical factor 1/$4{\pi}r^{2}$= 1.3${\times}10^{-46}{\rm cm}^{-2}$
 for r = 8 kpc = 2.5${\times}10^{22}$ cm. (We call this geometrical reduction factor  {\bf  $\sharp$H}.)

(6) Since the data of the COS-B satellite is given in the units of (cm$^{2}$ sec str)$^{-1}$ [15], we must calculate the solid angle of the bulge. While the solid angle of the Galactic center cannot be definitely determined, the bright region of the Galactic bulge spreads over approximately 5 degrees in the Galactic longitude direction and 2 degrees in Galactic latitude. Therefore, we estimate the field of view of the Bulge to be 5$\times$2 square degrees. This corresponds to an opening angle of 0.087 radians $\times$0.035 radians or 3.0${\times}10^{-3}$ steradians. The data of COS-B towards the Galactic bulge give a gamma-ray flux for the Galactic center of 8$\times10^{-4 }{\rm cm}^{-2}{\rm sec}^{-1}{\rm str}^{-1}$ at $E_{\gamma}{\geq}$70 MeV to 5 GeV for $\ell = 11-20^{\circ}$.   Multiplying the field of view by this number gives an intensity of gamma-rays per unit area from the direction of the Bulge of 2.4$\times10^{-6}{\rm cm}^{-2}{\rm sec}^{-1}$.  (We call this number  {\bf  $\sharp$I})

(7) We now compare the observed data with our predictions. According to our prediction, the flux of gamma-rays from the Galactic center, obtained by multiplying  {\bf  $\sharp$G} and 
 {\bf  $\sharp$H}, is (2.5$\times10^{40}$ photons/sec) $\times$ (1.3$\times10^{-46}{\rm cm}^{-2}$) = 3.3$\times10^{-6}$photons/cm$^{2}$sec at $E_{\gamma}{\geq}$70 MeV.   (We call this number  {\bf  $\sharp$J}.) The predicted number  {\bf  $\sharp$J} is 35 $\%$ 
higher than the observation data  {\bf  $\sharp$I}.   However, given the crude estimation process, the coincidence of the two numbers is surprising. This shows that it may be possible to determine the origin of cosmic rays by simple calculations of this type.

   We are aware that there are numerous points of possible improvement for future studies. For example, we need to determine if the value of 20 million years is reasonable for the time cosmic rays are retained in the Galaxy and whether the photon intensity be fairly approximated by the proton flux divided by a of factor 12.  

 Furthermore, we have not taken account of the contribution by primary electrons, which are expected to contribute to the flux of gamma-rays in the low energy region $E_{\gamma}$
$\leq$70\,MeV since electrons are simultaneously accelerated with protons in solar flares. 
Probably electrons will be accelerated by stellar flares upto $\sim$ 100\,MeV: the same 
Lorentz factor of protons.

We must also take into account flares induced by nearby stars, which are likely to be quite strong.   Additionally, we must consider the re-acceleration process of  {\it seed protons} produced by flares in super novae remnants.  By these processes protons will be re-accelerated into higher energies ( {\it the second step acceleration}) and will have a harder energy spectrum.  These protons should properly be classed as cosmic rays.   

\section{Summary and Conclusions}

 (1) The origin of the diffuse gamma-rays is likely to be protons accelerated by stellar flares
({\it the first step accleration}). They are not chiefly produced by super novae or the interstellar medium.
See Figure 2.

(2) The energy spectrum of primary protons may be expressed by  ${E_1}^{\Gamma}$ with the index $\Gamma$ = -3.75

(3) The accelerated protons by stellar flares are retained not only near the Galactic bulge but also in the Galactic disk and halo. They are retained in the Galaxy for approximately 20 million years.

(4) When these protons collide with the interstellar medium, gamma-rays are produced. These gamma-rays are likely to be the diffuse gamma-rays observed by the COS-B, EGRET and Fermi satellites 
around  ${E_{\gamma}{\approx}}$100 MeV [16-17].  For the production process, our model is based on the Hayakawa$-$Hutchinson$-$Morrison hypothesis.
However contribution of pp collisions would be not restricted down to 70 MeV and in the lower neergy region, probably the contribution of electrons would be dominated.

(5) The primary accelerated particles are re-accelerated in the interstellar shocks to the energy $E_{2}$.  The energy spectrum can be written by $E_{2}^{\Gamma}dE_{2}$and $\Gamma$=-2.75. 

(6) The observed flux of the diffuse gamma-rays with  $E_{\gamma}$$\geq$a few GeV must be produced 
by these re-accelerated protons.  However, the intensity  of diffuse gamma-rays with around ${E_{\gamma}{\sim}}$70MeV may be produced by primary accelerated protons by stellar flares.

\section{Acknowledgments}
The author wishes to acknowledge the contributions of the late professor Satio Hayakawa, who instructed him at Nagoya University and who demonstrated how interesting the study of cosmic rays could be.

%\begin{figure}[th]
\begin{figure}
\centering
\includegraphics[width=2.7in]{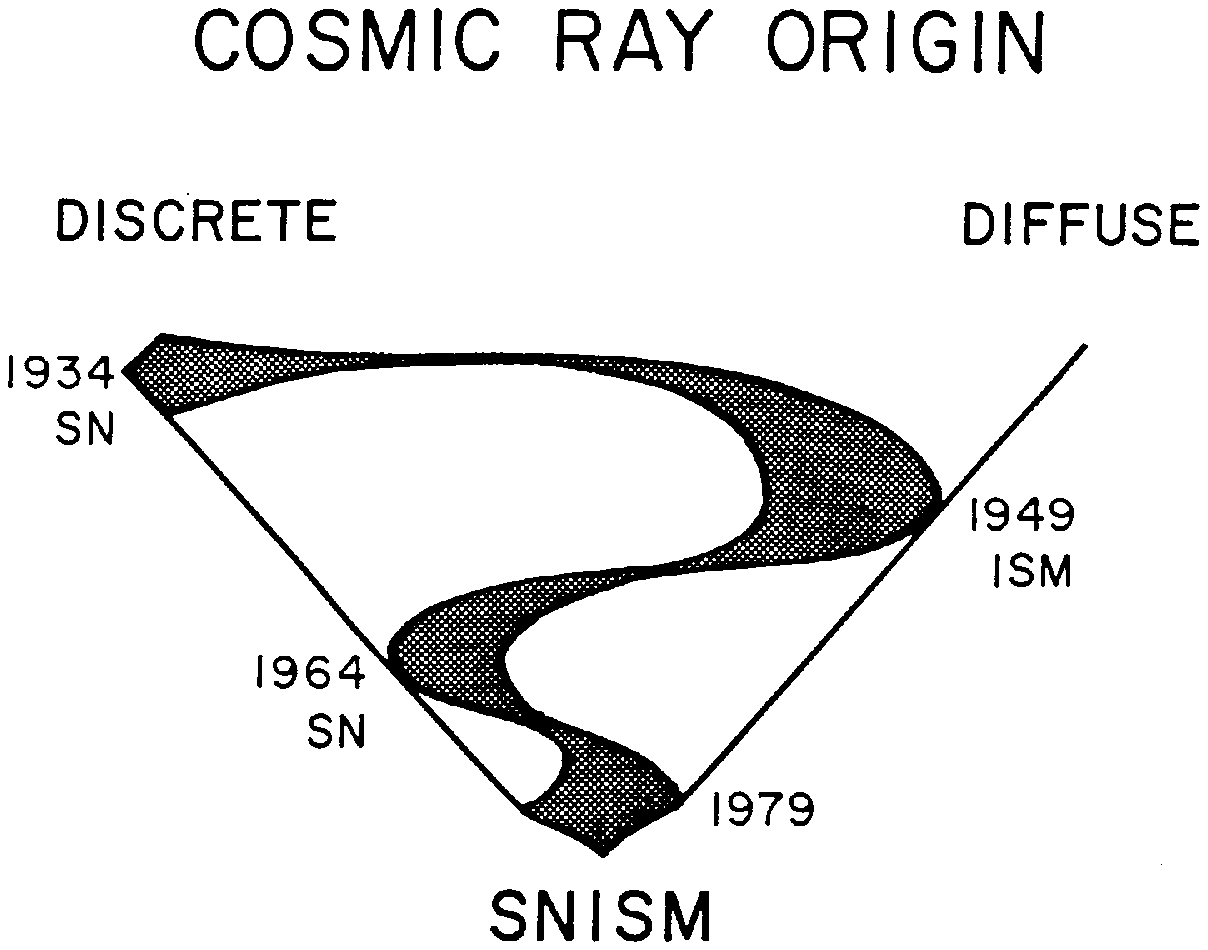}
\caption{A famous picture given by Lingenfelter on the origin of cosmic rays.  In very early days, the origin of
cosmic rays was considered as to be super nova, however after Fermi's idea, the origin of
cosmic rays was considered as interstellar matter.  At Kyoto conference in 1979, 
Lingenfelter had pointed out that the seed particles of cosmic rays should be both origin named SNISM supernova interstellar matter}
\label{fig1}
\end{figure}

\begin{figure*}
\centering
\includegraphics[width=6.5in]{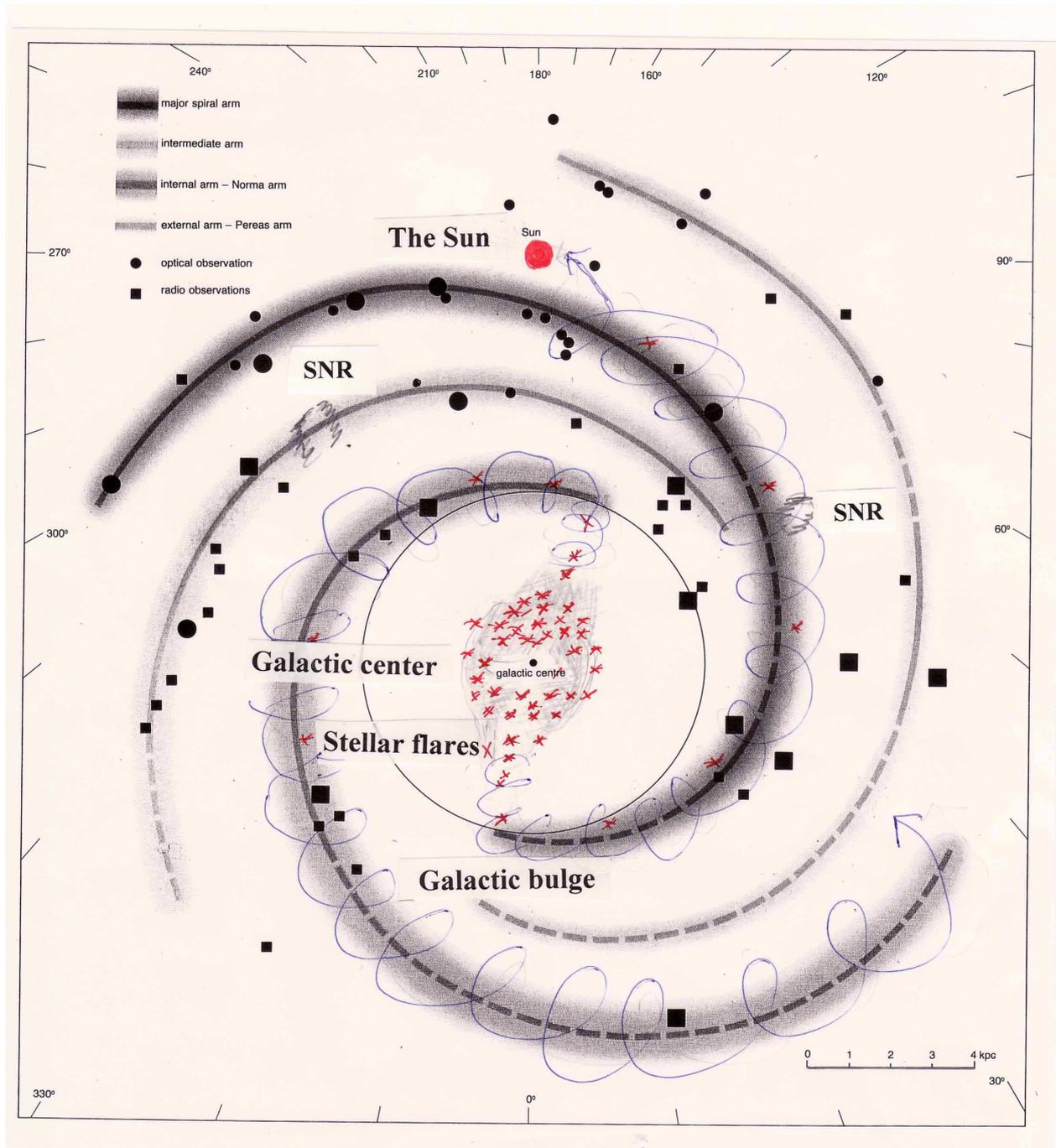}
\caption{A picture where seed particles of cosmic rays are produced.  The
seed particles of cosmic rays may be coming from stellar flares in the Galactic bulge and
arm.   The red cross marks represent where stellar flares occur.   Those accelerated
protons by stellar flares will be transported to the Sun through the Galactic arm.  
They will sometimes penetrate the region of super nova remnants.  There must be matter turbulance 
and/or strong magnetic field.   Then either by the original second Fermi acceleration mechanism or 
the first Fermi acceleration mechanism, these particles will be re-accelerated into high energies.
They will have harder energy spectrum than that of original accleerated protons.   These are 
called as cosmic rays.
The basic picture is given by the coutesy of Cambridge University Press.   
ISBN 0 521 26369 7 (1985)  The Cambridge Atlas of Astronmy, 
edited by J. Audouze and G. Israel. }
\label{fig2}
\end{figure*}


\begin{thebibliography}{99}
\bibitem{Matsubara}Y. Matsubara, private communication, According to him,
the frequency of the flares N can be written as the power of the size s with the slope -1.3
or ${N{\propto}s^{-1.3}}$.
\bibitem{Note} Note added:
Here we have assumed the inelasticity of gamma-rays to be $K_{\gamma}{\approx}1/12$.
  In proton$-$proton interactions in the interstellar medium, the accelerated protons loose half of the energy of the incoming proton on average (an inelasticity of $K {\approx} 0.5$). One sixth of the initial energy of the incoming proton is dissipated into the production of two charged pions ($\pi^{+}\pi^{-}$) and one neutral pion ($\pi^{o}$). The neutral pion decays into two photons, and hence the inelasticity of $K_{\gamma}$ can be expressed as ${\approx} 1/12 (= 1/2{\times}1/3{\times} 1/2$).   %Here we have assumed that
\end{thebibliography}
\end{document}